\documentclass[structabstract]{aa}  
\usepackage{graphicx}
\usepackage{txfonts}
\usepackage{natbib}
\usepackage{graphicx}
\usepackage{epstopdf}
\usepackage{ulem}

\begin{document}
   \title{Detection of OD towards the low-mass protostar IRAS 16293$-$2422}

   \subtitle{}

   \author{B. Parise
          \inst{1}
          \and
          F. Du \inst{1}
          \and
          F.-C. Liu \inst{1}
          \and
          A. Belloche \inst{1}
          \and 
          H. Wiesemeyer \inst{1}
          \and 
          R. G\"usten \inst{1}
          \and 
          K. M. Menten \inst{1}
          \and
          H-W H\"ubers \inst{2}
          \and 
          B. Klein \inst{1}
                    }

   \institute{Max-Planck-Institut f\"ur Radioastronomie,
              Auf dem H\"ugel 69, 53121 Bonn, Germany
             \\
              \email{bparise@mpifr-bonn.mpg.de}
         \and
             Deutsches Zentrum f\"ur Luft- und Raumfahrt, Institut f\"ur Planetenforschung,  Rutherfordstra{\ss}e 2, 12489 Berlin, Germany
             }

   \date{Received xxx ; accepted xxx }

  \abstract
   {Although water is an essential and widespread molecule in star-forming regions, its chemical formation pathways are still not very well constrained. Observing the level of deuterium fractionation of OH, a radical involved in the water chemical network, is a promising way to infer its chemical origin. }
   {We aim at understanding the formation mechanisms of water by investigating the origin of its deuterium fractionation. This can be achieved by observing the abundance of OD towards the low-mass protostar IRAS16293$-$2422, where the HDO distribution is already known.}
   {Using the GREAT receiver on board SOFIA, we observed the ground-state OD transition at 1391.5 GHz towards the low-mass protostar IRAS16293$-$2422.
   We also present the detection of the HDO 1$_{11}$-0$_{00}$ line using the APEX telescope. We compare the OD/HDO abundance ratio inferred from these observations with the predictions of chemical models. }
   {The OD line is detected in absorption towards the source continuum. This is the first detection of OD outside the solar system. The SOFIA observation, coupled to the observation of the HDO 1$_{11}$-0$_{00}$ line, provides an estimate of the abundance ratio OD/HDO $\sim$ 17--90 in the gas where the absorption takes place. This value is fairly high compared with model predictions. This may be reconciled if reprocessing in the gas by means of the dissociative recombination of H$_2$DO$^+$ further fractionates OH with respect to water.}
   {The present observation demonstrates the capability of the SOFIA/GREAT instrument to detect the ground transition of OD towards star-forming regions in a frequency range that was not accessible before. Dissociative recombination of H$_2$DO$^+$ may play an important role in setting a high OD abundance. Measuring the branching ratios of this reaction in the laboratory will be of great value for chemical models. }


   \maketitle

\section{Introduction}

Water is an essential molecule in star-forming regions because it is in certain conditions one of the main gas coolants \citep[e.g.][]{Nisini10}, 
and it is the most abundant constituent of the icy mantles frozen onto dust grains \citep[e.g.][]{Gibb04}. Despite this fame, water is still mostly a mystery for astrochemists: its formation pathways are not very well constrained. Water can form in the gas phase at high temperatures, or on the dust grains via surface chemistry. 
Surface chemistry in particular is believed to play an important role in the cold and dense clouds where stars form. But the details of these processes are mostly unknown. The reaction network involves hydrogenation of oxygen atoms (O) and O$_3$ to form OH, which can in turn react with H or H$_2$ to form H$_2$O, 
as well as successive hydrogenations of O$_2$, leading to HOOH and water \citep[e.g.][]{Tielens82}.
This reaction network is at the moment the subject of intense laboratory studies \citep[e.g.,][]{Oba09,Ioppolo10,Cuppen10}. Some new observational insights into the water formation pathways have been achieved with the first detection in space of hydrogen peroxide (HOOH) towards the $\rho$ Oph A cloud \citep{Bergman11b}. HOOH is the precursor of water in the hydrogenation of O$_2$ on grain surfaces. Detailed coupled gas-grain chemical modelling of this detection was able to simultaneously reproduce the observed abundances of formaldehyde, methanol, O$_2$ and hydrogen peroxide in this source \citep{Du12}, validating the importance of grain chemistry processes in the formation of HOOH and as a consequence in water formation. 

We propose here yet another, complementary, approach for the study of water formation.
The level of deuterium fractionation\footnote{Enrichment in deuterium in a molecule, compared with the cosmic D/H ratio \citep[$\sim$ 1.5 $\times$ 10$^{-5}$,][]{Linsky03}.} in molecules has been shown to be a good signpost of their formation conditions. Molecules forming in cold and dense environments are observed to show a high level of fractionation, 
which even led to the detection of triply-deuterated molecules such as ND$_3$ \citep{vanderTak02} and CD$_3$OH \citep{Parise04}. The observation of the deuterium fractionation in a molecule and its molecular precursors has proven to be very useful in constraining the relative importance of formation pathways. For example, the observation of formaldehyde (H$_2$CO) and isotopologues (HDCO and D$_2$CO) as well as methanol (CH$_3$OH) and isotopologues (CH$_2$DOH, etc.) in a sample of low-mass protostars has shown that methanol is mainly formed on dust surfaces by hydrogenation of CO, whereas a 
significant fraction of formaldehyde must be formed in the gas phase \citep{Parise06a}. Another promising example is the study of the NH and ND radicals with the aim to understand the formation of NH$_3$ and isotopologues \citep{Bacmann10}.  

\begin{figure}[!ht]
\begin{center}
\includegraphics[width=9cm,trim=1.5cm 0.3cm 0 1cm, clip=true]{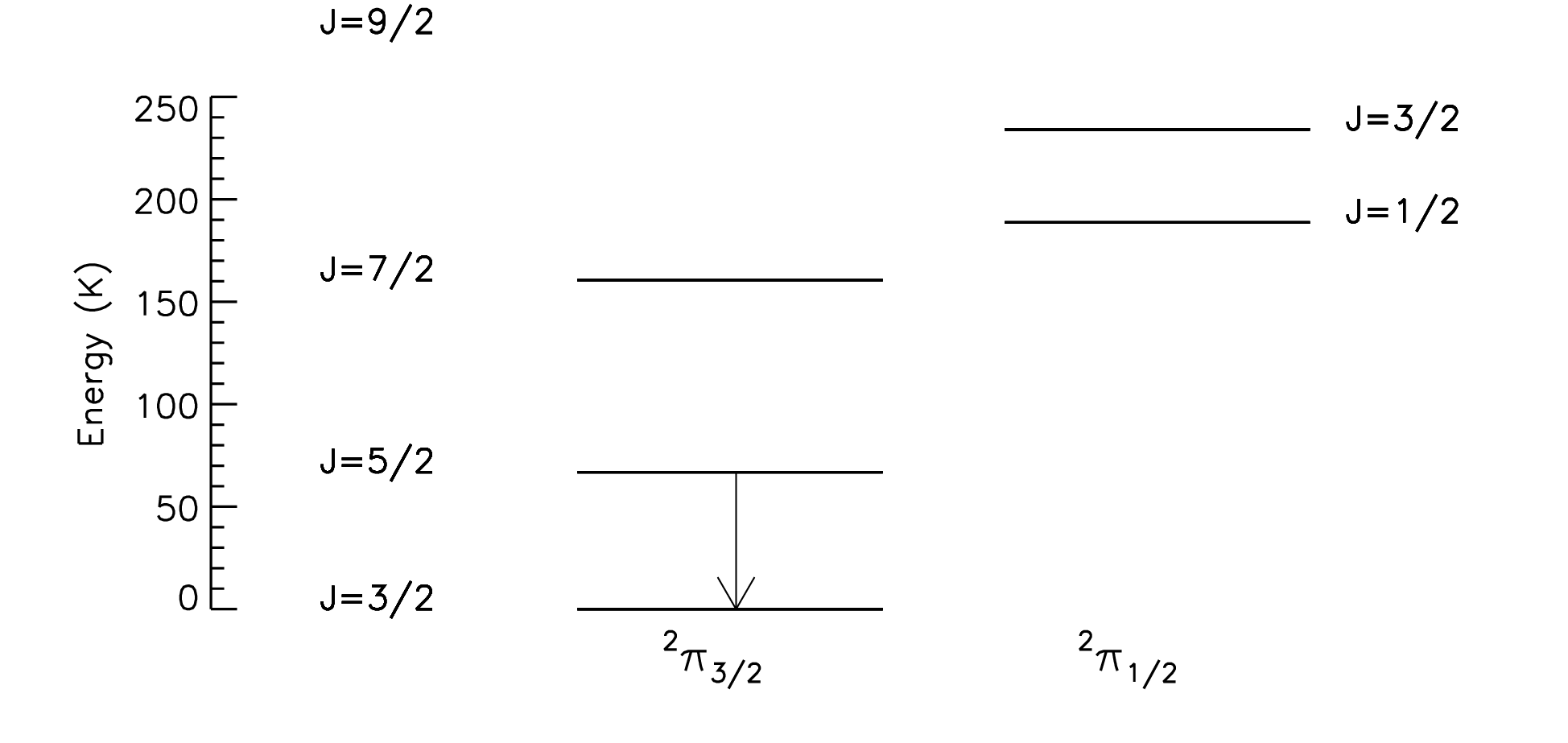}
\caption{Lowest energy levels of the OD radical. The arrow shows the observed line at 1391.5\,GHz. Each rotational level on this figure is split to first order into a $\Lambda$-type doublet ($l$ = $\pm$1), each $l$ level is then split again into hyperfine sublevels (see Fig. \ref{hyp}).}
\label{energy_od}
\end{center}
\vspace{-0.45cm}
\end{figure}
 
Water formation on dust surfaces is believed to proceed via the hydrogenation of either OH or HOOH, the latter case leading to the production of an additional OH. The abundance of H$_2$O and OH should be therefore tightly linked if grain processes are dominant. The deuterium fractionation of water in hot corinos (i.e. in the gas enriched from evaporation of dust mantles in low-mass protostars) is found to be anomalously low in comparison with other molecules \citep[e.g.,][]{Parise05ashort, Liu11}, and the reason for this is not yet understood. Therefore, observations of the OD/OH ratio in these same regions will help to understand the peculiarity of water chemistry in young stellar objects. In the present study, we target OD towards the low-mass protostar IRAS16293$-$2422 (hereafter IRAS16293), where the distribution
of deuterated water has been previously studied \citep{Parise05ashort}.
We note that the OD/OH ratio is predicted to be relatively high if it is formed exclusively in the gas phase \citep[more than $\sim$\,0.1 at $T$\,=\,10\,K and $n_H$\,=\,10$^4$\,cm$^{-3}$,][]{Roberts02c}. 

 OD has been searched to date unsuccessfully towards the Galactic Centre, by means of the observation of the $\Delta$F = 0 $\Lambda$-doublet transitions of the $^2\Pi_{3/2}$, J=3/2 state around 310 MHz \citep{Allen74}. This non-detection may now be explained by the lower D/H ratios observed in general towards the Galactic Centre, due to astration (destruction of deuteron nuclei in stellar interiors). OD has been recently
 detected in the comet C/2002 T7 (LINEAR) via the co-addition of 30 lines of its ultraviolet fluorescence spectrum, with a ratio [OD]/[OH] = (3.5$\pm$1.0)\,$\times$\,10$^{-4}$ \citep{Hutsemekers08}.
 These observations, interpreted as the result of photodissociation of HDO and H$_2$O, lead to an estimate of the water fractionation of the same order as in other comets, a factor of 10 lower than observed in hot corinos. 
  
The present paper is organised as follows. Section \ref{spec} briefly describes the spectroscopy of OD. Section \ref{obs} presents the observations, which are analysed in Sect. \ref{analysis}. Results are discussed in Sect. \ref{results}.

\section{OD spectroscopy}
\label{spec}

\begin{figure*}[!h]
\begin{center}
\includegraphics[trim=1.3cm 3cm 0 2.0cm, clip=true,width=13cm]{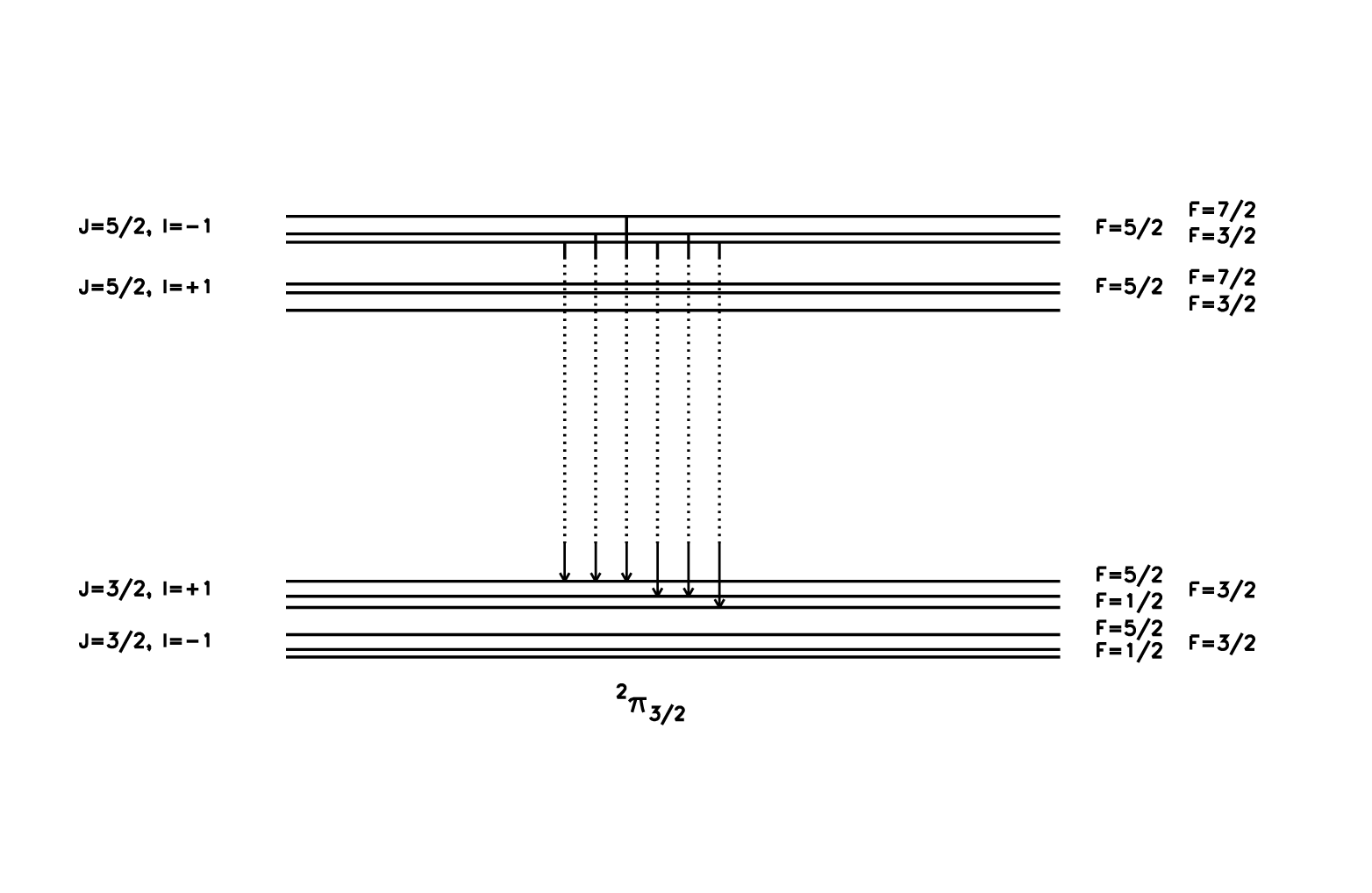}
\caption{Hyperfine structure of the 1391.5\,GHz transition. The spacing between the different hyperfine levels was increased for visibility, and the overall separation between the J=5/2 and 3/2 levels is not to scale. The transitions J=5/2$\rightarrow$3/2, l=+1$\rightarrow$$-$1 at 1390.6\,GHz, difficult to observe from SOFIA because of residual atmospheric absorption, are not shown. }
\label{hyp}
\end{center}
\end{figure*}

Figure \ref{energy_od} shows the lower energy levels of the OD molecule.
Owing to the unbalanced electronic spin and orbital angular momenta, the spectrum of the OD radical shows $\Lambda$-type doubling, similar to that of the OH radical. Each rotational level is split into a $\Lambda$-type doublet. Each component of the doublet is further split by the magnetic hyperfine structure (hfs) due to the deuterium atom into components characterised by the total angular momentum F (=J+I). F takes values of J$-$1, J and J+1 \citep{Dousmanis55}. The selection rule of $\Delta$F=0,$\pm$1 leads to six hfs lines (Fig. \ref{hyp}).
We use here the transitions tabulated in the JPL database\footnote{http://spec.jpl.nasa.gov/} \citep{Pickett98}. These predictions are based on the laboratory measurements of \citet{Beaudet78} and \citet{Brown82}.

\section{Observations}
\label{obs}

\begin{figure}[!h]
\centering
\includegraphics[width=9cm,trim= 0cm 0cm 0cm 0.8cm,clip=true]{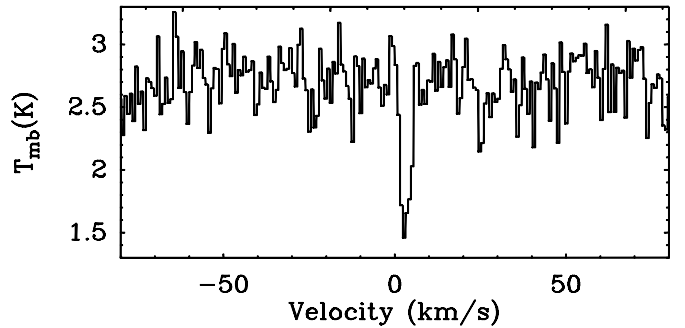}
\caption{OD J=5/2$\rightarrow$3/2, l=--1$\rightarrow$+1 transition observed towards IRAS16293. The reference frequency is that of the main hfs component. The spectrum is smoothed to 0.79\,km\,s$^{-1}$ resolution. The continuum level is twice the true continuum because of DSB calibration. }
\label{spectrum_od}
\end{figure}

Using the GREAT instrument\footnote{GREAT is a development by the MPI f\"ur Radioastronomie and the KOSMA/Universit\"at zu K\"oln, in cooperation with the MPI f\"ur Sonnensystemforschung and the DLR Institut f\"ur Planetenforschung.} \citep{Heyminck12} on board SOFIA, we observed the OD $^2\Pi_{3/2}$ J=5/2\,$\rightarrow$\,3/2, l=$-$1\,$\rightarrow$\,+1 line multiplet towards the low-mass protostar IRAS16293 at the coordinates $\alpha$(2000)=16$^{\rm h}$32$^{\rm m}$22\fs90, $\delta$(2000)=$-$24$^\circ$28$'$36\farcs3. The double-sideband (DSB) receiver was tuned in the upper sideband at the frequency 1391494.71 MHz of the strongest hfs component. The receiver was connected to a digital FFT spectrometer \citep{Klein12} providing a spectral resolution of 0.04\,km\,s$^{-1}$,  
and a bandwidth of 1.5\,GHz. Two setups were observed with different LSR velocities ($-$2\,km\,s$^{-1}$ and +8\,km\,s$^{-1}$) to disentangle any contamination from the image sideband.
We used the double-beam switch mode chopping in E-W direction with an
amplitude of 70$''$ and a phase time of 0.5 sec. The atmospheric transmission was fit with special care during the calibration process to remove a residual telluric line at 30 km/s.
Part of the data suffer from a pointing error in the second phase of our double-beam observations. 
Because the OD line is observed in absorption against the continuum of the source, which is likely compact at this frequency 
compared with the SOFIA beam (20\farcs4 at the OD frequency), a pointing offset should only result
in a different coupling of the signal with the telescope beam. We therefore scaled all ON-OFF cycles to the brightest continuum level and averaged them. The resulting spectrum is obtained with 11 min on-source integration time. An additional standing-wave in the signal was removed by killing the appropriate peak at $\sim$\,0.05 MHz$^{-1}$ in the Fourier plane of the signal. The resulting spectrum is presented in 
Fig. \ref{spectrum_od}. The beam efficiency is taken to be 0.54 \citep[]{Heyminck12}.  
The OD feature clearly appears in absorption, with a peak signal-to-noise ratio of 6 at the resolution of 0.79\,km\,s$^{-1}$, the rest of the band being free of any emission or absorption. The DSB-continuum is 2.7$\pm$0.2 K at the same resolution.

\begin{figure}[!ht]
\centering
\includegraphics[width=8.0cm,trim= 0cm 0cm 0cm 0.8cm,clip=true]{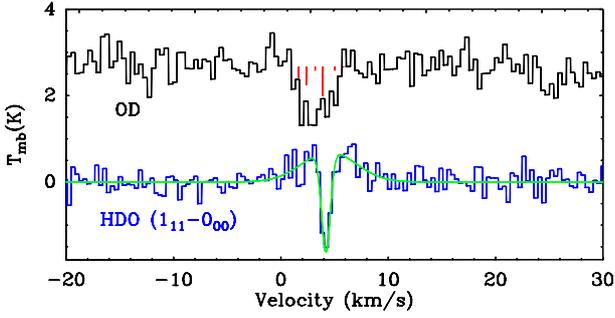}
\caption{HDO\,1$_{1,1}$--0$_{0,0}$ and OD line profiles observed towards IRAS16293. The HDO spectrum was shifted by -2.9\,K along the y-axis.
The OD spectrum was smoothed to 0.39\,km\,s$^{-1}$. The hyperfine intensity structure of the OD line is shown as red bars. Note that the OD observation is DSB while the HDO observation is SSB. The green curve shows the two-Gaussian fit of the HDO profile (see Table \ref{tab}).} 
\label{spectrum}
\vspace{-0.3cm}
\end{figure}

We also present here observations of the HDO 1$_{1,1}$--0$_{0,0}$ line obtained with the CHAMP$^+$ receiver at APEX. The observations were performed on 2010, September 6 and 7, under very stable atmospheric conditions (PWV\,=\,0.35\,$-$\,0.40 mm). The CHAMP$^+$ high-frequency channel was tuned to 893.639\,GHz, while the low-frequency channel was tuned on CO(6-5).
The focus was checked on Venus, and the local pointing was checked every 1 to 1.5 hours with CO(6-5) line pointing on NGC6334I. 
Pointing corrections were found to be of the order of 3$''$. The angular resolution of the HDO observations is 7$''$.
The receiver was connected to the AFFTS, providing a resolution of 0.06\,km\,s$^{-1}$, and a bandwidth of 1.5\,GHz.
The continuum is 2.9$\pm$0.2\,K (T$_{mb}$ scale) at resolution 0.37\,km\,s$^{-1}$.

\section{Analysis}
\label{analysis}

Figure \ref{spectrum} presents the OD spectrum observed with SOFIA (black curve), as well as the HDO line observed towards the same source with APEX. The OD spectrum is characterised by a broad absorption (2.9\,km\,s$^{-1}$, see Table \ref{tab}), compared with the HDO line (1.0\,km\,s$^{-1}$). This 
broadening is due to the hyperfine structure of the OD line. An hfs fit using the CLASS software\footnote{http://iram.fr/IRAMFR/GILDAS/} provides an intrinsic linewidth of 1.3$\pm$0.6 km\,s$^{-1}$, which is consistent with the width of the HDO absorption within the error bars.

\renewcommand{\tabcolsep}{1.5mm}
\begin{table}[!h]
\caption{\label{tab}Fitting results.}
\centering
\vspace{-0.2cm}
\begin{small}
\begin{tabular}{llccc}
\hline
\hline
\noalign{\smallskip}
Molecular line   &  Fit  type &    $\int$\,T$_{mb}$\,dv & FWHM & v$_{lsr}$\\
& & (K\,km\,s$^{-1}$) & (km\,s$^{-1}$)& (km\,s$^{-1}$) \\
\noalign{\smallskip}
\hline
\noalign{\smallskip}
HDO  1$_{1,1}$\,--\,0$_{0,0}$ &  two-Gauss & +4.4$\pm$0.6  &  5.9$\pm$1.0  & 4.5$\pm$0.3\\
           & &      $-$2.4$\pm$0.3 & 1.0$\pm$0.1  & 4.2$\pm$0.1\\
\noalign{\smallskip}
\hline
\noalign{\smallskip}           
OD ~{\tiny 5/2$\rightarrow$3/2, --1$\rightarrow$+1} & one-Gauss & $-$4.1$\pm$0.5 & 2.9$\pm$0.3 &-- \\
\noalign{\smallskip}
\hline
\noalign{\smallskip}       
OD ~{\tiny 5/2$\rightarrow$3/2, --1$\rightarrow$+1} & hfs &  --  & 1.3$\pm$0.6 & 4.2$\pm$0.2\\
\noalign{\smallskip}
\hline
\noalign{\smallskip}
\end{tabular}
\end{small}
\vspace{-0.3cm}
\end{table}

Collision rate coefficients for the OD molecule are not available. However, an estimate of the critical density of the upper level of the transition can be obtained 
by looking at the critical density of the corresponding line of OH (at 2514 GHz). Using the values of the Einstein coefficient for spontaneous emission
and the collision rates tabulated  in the LAMDA database\footnote{http://www.strw.leidenuniv.nl/$\sim$moldata/}, we find that the critical density
of the corresponding OH upper level is abound 10$^{10}$ cm$^{-3}$ at 60\,K. The critical density of the 894\,GHz HDO line has a similar value in 
the 50\,--\,100\,K range. We can therefore reasonably assume that both absorptions occur in the same gas, and that the excitation temperature of the lines 
will be extremely low, as the density of the IRAS16293 envelope is several orders of magnitude lower than this critical value.

We assume that each hyperfine component of OD has an intrinsic line width similar to that of the HDO absorption (1\,km\,s$^{-1}$), and we model
the global absorption using the XCLASS software\footnote{http://www.astro.uni-koeln.de/projects/schilke/XCLASS}. 
The OD line is close to saturation, as already visible in Fig.\,\ref{spectrum}, where the depth of the absorption is about half that of the DSB continuum (i.e. the line absorbs the totality of the continuum in the signal sideband, leaving the image sideband continuum unabsorbed). The saturation remains limited ($\tau$ $<$ 4), as no line broadening (except for that due to hyperfine structure) is observed. The signal-to-noise ratio of the observation is not high enough, however, to infer the optical depth 
from the comparison of the depth of the three main hyperfine components. 
The OD and HDO column densities inferred from the XCLASS modelling averaged over the extent of their respective continuum for different assumed $T_{ex}$ are tabulated in Table\,\ref{tab2}. The uncertainties are estimated by comparing by eye the observation with the synthetic model.
Because both lines are seen in absorption against the source continuum, and the absorbing layer is likely to be more extended than the continuum in each case,
we can compare the two column densities without beam dilution hindrance. 
With these assumptions, the abundance ratio OD/HDO is found to be in the range 17\,--\,90, if we assume that the OD and HDO absorptions come from the same region of the envelope, and are described by the same excitation temperature, uniform along the line-of-sight.

\begin{table}[!h]
\caption{\label{tab2}Column densities inferred from the XCLASS modelling.}
\centering
\vspace{-0.2cm}
\begin{small}
\begin{tabular}{lccc}
\hline
\hline
\noalign{\smallskip}
$T_{ex}$ (K) &  $N_{\rm OD}$ (cm$^{-2}$)  &    $N_{\rm HDO}$ (cm$^{-2}$)  &   OD/HDO \\
\noalign{\smallskip}
\hline
\noalign{\smallskip}
   2.7   &   (3.5$\pm$1.5)\,$\times$\,10$^{13}$  &  (6.0$\pm$1.5)\,$\times$\,10$^{11}$ & 60\,$\pm$\,30\\
   5.0   &   (5.0$\pm$2.0)\,$\times$\,10$^{13}$  &  (1.2$\pm$0.3)\,$\times$\,10$^{12}$ & 45\,$\pm$\,20\\
   10.0 &   (1.0$\pm$0.3)\,$\times$\,10$^{14}$  &  (4.0$\pm$1.0)\,$\times$\,10$^{12}$ & 27\,$\pm$\,10\\  
\noalign{\smallskip}
\hline
\noalign{\smallskip}       
\end{tabular}
\end{small}
\vspace{-0.7cm}
\end{table}

\section{Discussion}
\label{results}

Ideally, we would like to measure the OD/OH ratio in the envelope of IRAS16293 to compare it on the one hand with the HDO/H$_2$O ratio in the same environment, with the aim to constrain astrochemical models describing the synthesis of water and its deuterium fractionation, and on the other hand with the OD/OH ratio in more evolved environments, e.g. in comets, to infer the amount of chemical reprocessing during the formation of a solar-type star. 
To our knowledge, no column density for OH in the absorbing envelope of IRAS16293 has been reported so far. Although Herschel/HIFI could provide 
a high-resolution observation of the OH $^2\Pi_{1/2}$ J=3/2 $\rightarrow$ 1/2 line at 1.83\,THz, this line has an upper energy level too high to be directly compared to the OD line presented here.
A more promising perspective in the future 
will be to observe the corresponding OH line ($^2\Pi_{3/2}$ J=5/2\,$\rightarrow$\,3/2) at 2.5\,THz with the upgraded GREAT instrument. 

In the meantime, we discuss here the value of the OD/HDO column density ratio inferred in Sect. \ref{analysis}. In the absence of a differential fractionation
in OH and water, one expects the OD/HDO ratio to be close to the OH/H$_2$O abundance ratio.

\begin{figure*}[!ht]
\centering
\begin{tabular}{cc}
\includegraphics[trim=0cm 0 0 1.0cm, clip=true,width=8.0cm]{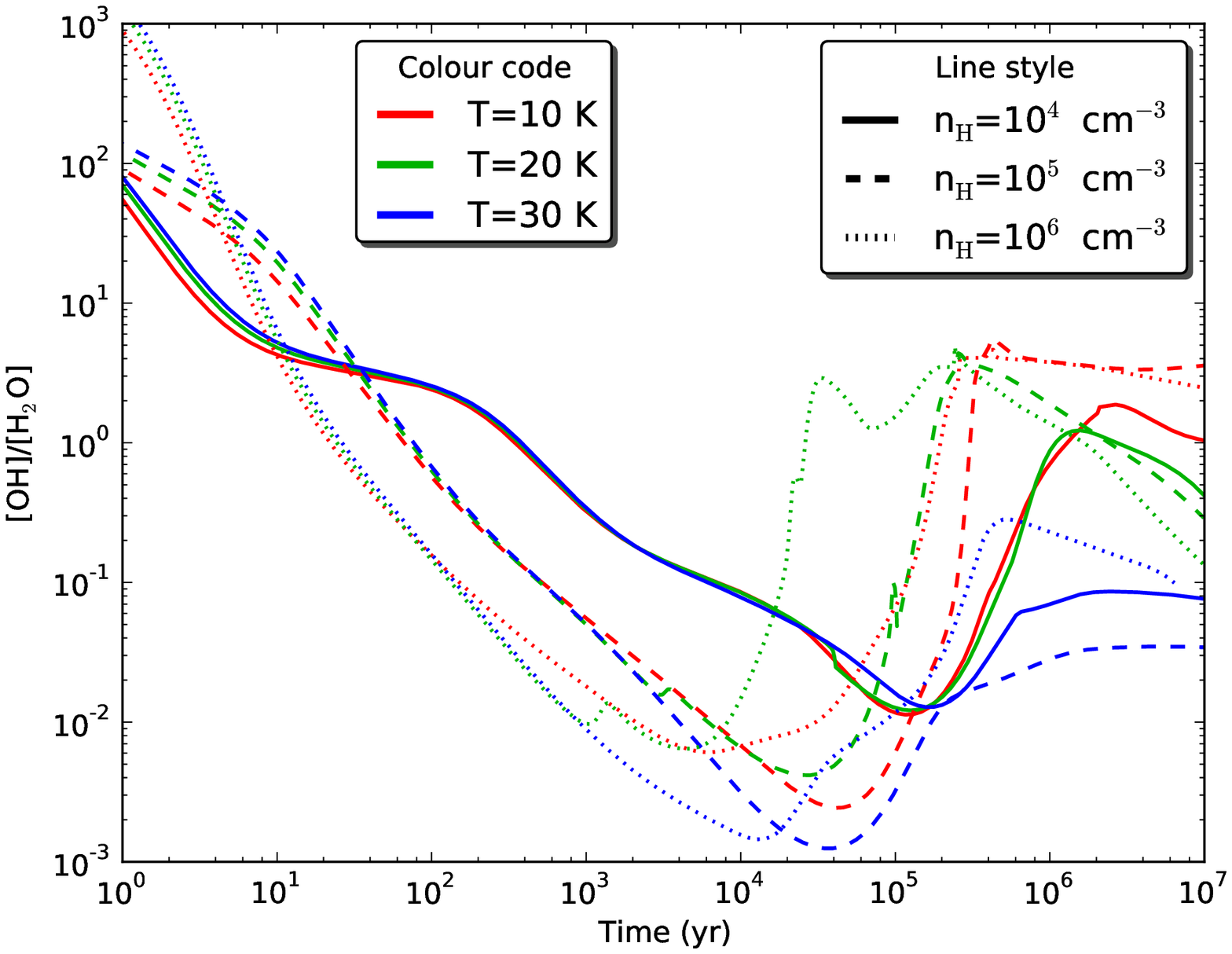} &
\includegraphics[trim=0cm 0 0 1.0cm, clip=true,width=8.0cm]{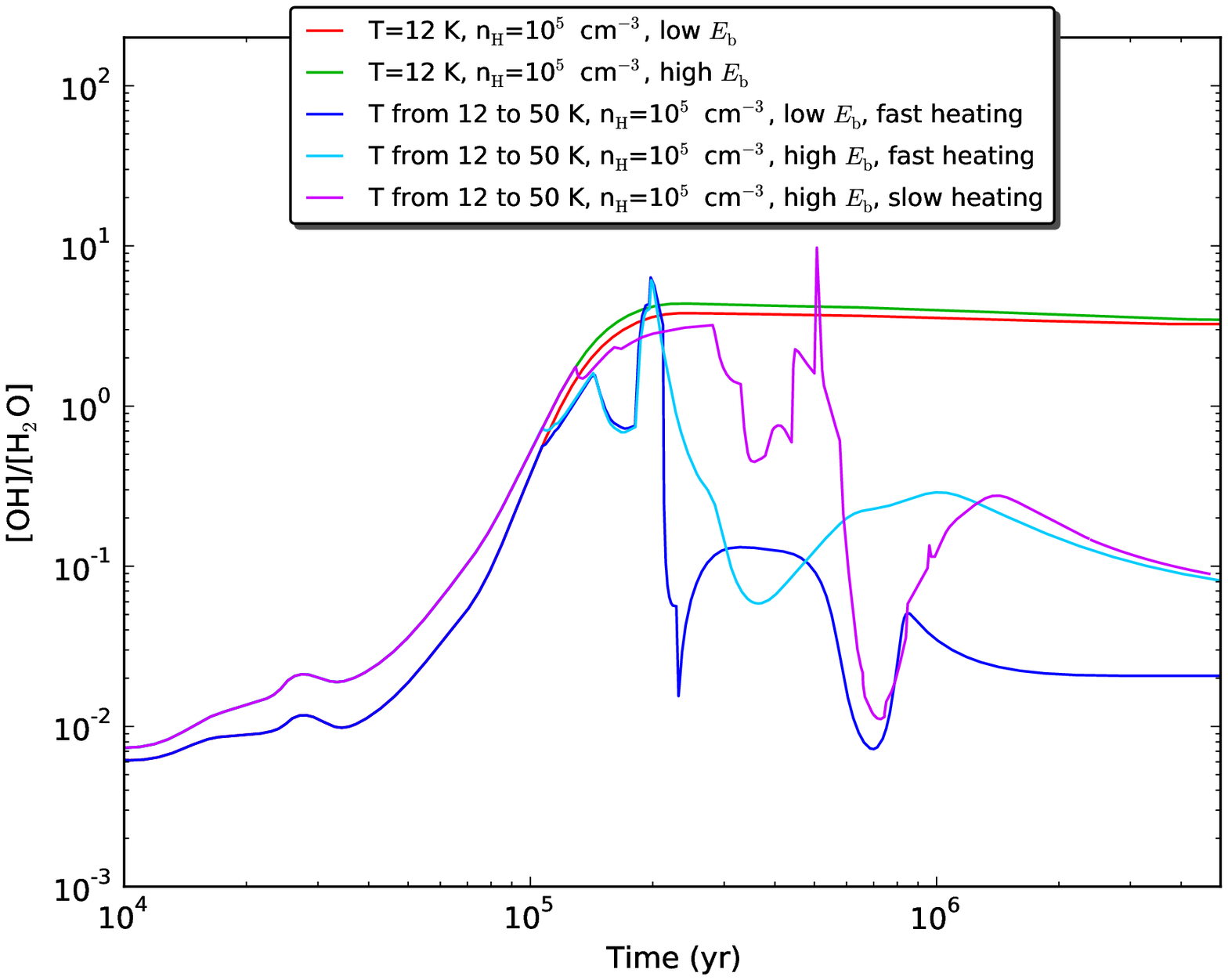} \\
\end{tabular}
\caption{Left panel: Predictions of the OH/H$_2$O ratio from the \citet{Du12} standard model, for constant ($T$, $n_H$) conditions. Right panel: Models with constant and time-dependent temperature.  ``Low $E_b$" models use the binding energies for species on the grains of the standard model of  \citet{Du12} while ``high $E_b$" models use the higher binding energies for species with an OH bond from \citet{Garrod06a}.}
\label{model}
\vspace{-0.3cm}
\end{figure*}

\citet{Du12} have presented a chemical model that includes processes in the gas-phase and on the grains, to describe the formation of water. We use the same model here, considering constant physical conditions, in the range $T$\,=\,10\,--\,30\,K and $n_H$\,=\,10$^4$\,--\,10$^6$\,cm$^{-3}$ (see Fig.  \ref{model}, left panel). Except at the very early times (t\,$<$\,10\,yr), the gas-phase 
OH/H$_2$O ratio is always predicted to be fairly low compared with our observed OD/HDO ratio. The closest agreement
happens for times later than 10$^4$\,--\,10$^5$\,yr, when the OH/H$_2$O can reach a maximum value of 5.7.

The models stated above assume constant temperature and density conditions. However, during the formation of a star,
the envelope is heated by the central protostar, and a model where the temperature evolves with time may be more appropriate. Figure \ref{model} (right panel)  
shows some results with a time-dependent temperature based on the model of \citet{Du12}. For the models with protostellar heating, $T$ increases between 12\,K and 50\,K linearly with time, either in the 1--3 10$^5$ yr (``fast'' heating) or in the 1--10 10$^5$ yr range (``slow'' heating). We also test different sets for the desorption energies of species from the  grains.
None of the models seem to be able to explain the high inferred OD/HDO ratio. There could be two reasons for this discrepancy: either the radiative transfer we used is too simplistic (in which case OD collisional rates are highly needed to go beyond this treatment), or the OH radical gets much more fractionated than water. This latter case may 
arise because the OH + D exchange reaction is exothermic \citep{Roberts00a}. The relative fractionation of OH with respect to H$_2$O could be even more enhanced if the gas-phase dissociative recombinations (DR) of H$_3$O$^+$ and H$_2$DO$^+$ have non-trivial branching ratios for the formation of water and hydroxyl.
Our model based on the UMIST06 rate assumes a produced OH:H$_2$O ratio of 3:1, consistent with the work of \citet{Jensen00}, who studied the DR of all deuterated isotopologues of H$_3$O$^+$ but H$_2$DO$^+$. Their results concerning the DR of HD$_2$O$^+$ shows that both HDO and D$_2$O are produced in similar amounts, whereas the OD:OH ratio is around 3.4, which may explain our high observed OD/HDO ratio. Experiments on the DR of H$_2$DO$^+$ are needed, however, before drawing a firm conclusion.

\section{Conclusions}

This Letter presents the first detection of OD outside the solar system, and demonstrates the capability of the SOFIA/GREAT instrument for breakthrough observations of spectral lines at high spectral resolution in a frequency band previously not covered (the OD line presented here falls into the frequency gap of the HIFI
receiver on board {\it Herschel}). The high inferred OD/HDO ratio in the envelope may be the signpost of gas-phase reprocessing through dissociative recombination of H$_2$DO$^+$. After the commissioning of the GREAT 2.5 THz (M-)channel,  parallel observations of the OD and the OH ground-state transitions could be performed routinely towards, a.o., star-forming regions. The observation of these radicals is expected to provide valuable constraints on the formation and fractionation of water. 

\begin{acknowledgements}
This article is dedicated to my dear daughter Marion.
   Based in part on observations made with the NASA/DLR Stratospheric Observatory for Infrared Astronomy. SOFIA Science Mission Operations is operated by the Universities Space Research Association, Inc., under NASA contract NAS2-97001, and the Deutsches SOFIA Institut under DLR contract 50 OK 0901. We are thankful to the ground and flight crews of SOFIA for their outstanding support, and the GREAT instrument team for performing these observations. BP, FD and F.-C.L are supported by the German Deutsche Forschungsgemeinschaft, DFG Emmy Noether
    project number PA1692/1-1.
\end{acknowledgements}

\bibliography{/Users/bparise/These/Manuscrit/biblio}
\bibliographystyle{aa}

\newpage
\section{Online material}

\end{document}